\documentclass[prd,preprint,nofootinbib]{revtex4}

\usepackage{graphicx}
\usepackage{amsfonts}
\usepackage{latexsym}
\usepackage{amsmath}
\usepackage{amssymb}
\usepackage{slashed}
\usepackage{epsfig}

\begin{document}

\title{Stochastic gravitational wave spectrum from cosmic string
 emitting gauge bosons and Majorana fermions}

\author{Nobuchika Okada}
 \email{okadan@ua.edu}
 \affiliation{Department of Physics and Astronomy, University of Alabama, Tuscaloosa, Alabama 35487, USA}

\author{Osamu Seto}
 \email{seto@particle.sci.hokudai.ac.jp}
 \affiliation{Department of Physics, Hokkaido University, Sapporo 060-0810, Japan}

%

\begin{abstract}
The effect of particle radiation on the spectrum of the stochastic gravitational
 wave background (SGWB) from cosmic strings is studied.
We consider cosmic strings in an Abelian-Higgs model coupling with Majorana
 fermion whose mass is generated by the Higgs field, motivated by 
 a gauged $U(1)_{B-L}$ model, in which the Majorana fermions are identified with right-handed neutrinos.
Taking the energy loss by particle radiation into account,
 we evaluate the resultant SGWB spectrum and demonstrate the emergence of
 very high frequency cutoff due to the particle radiation.
\end{abstract}

\preprint{EPHOU-26-008} 

\vspace*{3cm}
\maketitle


\section{Introduction}

Cosmic strings are one-dimensional topological defects
 at nontrivial vacua appeared by a spontaneous symmetry breaking 
 and formed by the Kibble mechanism~\cite{Kibble:1976sj} at the phase transition
 in the early Universe~\cite{Hindmarsh:1994re,Vilenkin:2000jqa}. 
Those cosmic strings underwent non-trivial dynamics and form cosmic string network
 consist of long (infinite) string and string loops.
String loops are generated by the self-interesection of long string
 or reconnection at a collision of two strings.
Gravitational waves (GWs) radiation is one main mode of the energy loss
 of the cosmic string loops and would be detected as the stochastic GWs
 background (SGWB)~\cite{Vilenkin:1981bx,Vachaspati:1984gt,Hogan:1984is}. 
For a recent review, see for instance Refs.~\cite{Auclair:2019wcv,Schmitz:2024gds}.
The SGWB spectrum generated by cosmic string is very unique~\cite{Vilenkin:2000jqa,Caldwell:1991jj}.
It is damped at low frequency and is flat at high frequency.
This transition is caused by the change of background spacetime evolution,
 namely, from radiation dominated to matter dominated Universe. 

Srednicki and Theisen~\cite{Srednicki:1986xg}
 and Brandenberger~\cite{Brandenberger:1986vj} initiated to estimate a non-gravitational energy loss
 of cosmic strings through the particle emission.
A phase transition forming cosmic strings would be associated with the gauge symmetry breaking
 by Higgs mechanism, and the string configuration in such an Abelian-Higgs (AH) model is
 described by the so-called Nielson-Olsen solution~\cite{Abrikosov:1956sx,Nielsen:1973cs}. 
The emission of the gauge bosons was also estimated~\cite{Vilenkin:1986zz,Spergel:1986uu,Copeland:1987yv,Blanco-Pillado:2000nbp}.
Cosmological implication of particle emission for
 baryogenesis~\cite{Kawasaki:1987vg,Mohazzab:1994xj,Jeannerot:1996yi} and
 dark matter production~\cite{Jeannerot:1999yn,Cui:2007js,Cui:2008bd,Hindmarsh:2013jha,Long:2019lwl,Kitajima:2022lre,Kitajima:2023vre} as well as
 the emission of cosmic rays~\cite{MacGibbon:1989kk,Bhattacharjee:1989vu,Gill:1994ic,Wichoski:1998kh,Berezinsky:2011cp,Creque-Sarbinowski:2022mex} have been investigated.
Vachaspati and his collaborators have been developing the formulation of
 particle emission from cosmic strings~\cite{Long:2014mxa,Matsunami:2019fss}.

We study cosmic strings losing the energy by emitting
 gravitational radiation as well as particles,
 and evaluate the spectrum of the SGWB from cosmic strings
 with taking particle radiation into account.
Kitajima and Nakayama~\cite{Kitajima:2022lre,Kitajima:2023vre} as well
 as Cheng and Visinelli~\cite{Cheng:2024axj}
 evaluated SGWB spectrum from cosmic string emitting very light gauge boson or dark photon
 based on a dark $U(1)$ gauge theory.
Auclair, Steer and Vachaspati also calculated SGWB spectrum from cosmic strings
 with the particle emission rate being parameterized without specifying
 particle physics Lagrangian~\cite{Auclair:2019jip}.
In wide class of models, a Higgs field does not only generate the mass of
 the gauge boson but also often gives the mass for fermion $\psi$.
Such a well-motivated example from the viewpoint of particle physics
 is the gauged $U(1)_{B-L}$ model~\cite{Pati:1973uk,Davidson:1978pm,Mohapatra:1980qe,Mohapatra:1980} with the charge of the Higgs field being $-2$,
 where the extra neutral gauge boson $Z'$ and right-handed neutrino acquire
 the mass from its vacuum expectation value (VEV), and neutrino mass can
 be explained by the type-I seesaw mechanism~\cite{SeesawM,SeesawY,SeesawG,SeesawMS}.
In this paper, we estimate the spectrum of the SGWB from cosmic strings
 which also emit massive $Z'$ gauge boson and Majorana fermion $\psi$.

This paper is organized as follows. 
In Sec.~II, we give a brief review on the model of SGWB adopted in our analysis.
In Sec.~\ref{sec:emissionrate} we estimate the emission rate for the particles
 of the massive gauge boson $Z'$ and Majorana fermion $\psi$ from strings.
In Sec.~\ref{sec:spectrum}, after we solve the evolution of cosmic string length
 by the energy loss due to gravitational as well as particle radiation,
 we present the spectrum of SGWB for several benchmark points.
Section~\ref{sec:summary} is devoted to our summary.

\section{Gravitational radiation} 

The differential energy density of gravitational radiations with respect to the angular
 frequency $\omega$ emitted at time $t_1$, the emitted angular frequency $\omega_1$
 and observed at time $t$ is expressed as~\cite{Vilenkin:2000jqa} 
\begin{equation}
\frac{d \rho_\mathrm{gw}}{d\omega}(t) = G \mu^2 \int_{t_*}^t dt_1 \left(\frac{a(t_1)}{a(t)}\right)^3
 \sum_n \mathtt{n}(\ell_n, t_1) P_n,
\label{eq:drhog_domega}
\end{equation}
where $G$ is the gravitational constant, $\mu$ is the energy per unit length of a string,
 $\mathtt{n}(\ell, t)$ is the number density of non-self intersecting string loops with the invariant length $\ell$
 at the time $t$.  $P_n$ is the averaged power of emitted GWs by the $n$-th harmonic of loop oscillations with the frequency $f_n$. 
The (angular) frequency redshifts as 
\begin{equation}
 \omega = \frac{a(t_1)}{a(t)}\omega_1 ,
\end{equation}
 as the scale factor $a(t)$ increases, and $t_*$ is the time when the rapid motion of strings has damped.
The decay constant of the strings by the gravitational radiation $\Gamma$, defined by
\begin{align}
\dot{E} = \Gamma G \mu^2 ,
\label{eq:Power}
\end{align}
with $E$ being the emitted gravitational radiation energy from string loops, is decomposed as
\begin{align}
\Gamma = \sum_{n=1}^{\infty} P_n .
\end{align}
In the following, we consider only the emission form cusps and ignore kink contributions for simplicity.  In this paper, we take $\Gamma =50$.

The present differential density parameter of SGWB produced by cosmic strings  with respect to the frequency $f$ is given by
\begin{equation}
\Omega_\mathrm{gw}(f) = \frac{f}{\rho_\mathrm{cr}}\frac{d \rho_\mathrm{gw}}{d f}  = 8\pi(G\mu)^2 \frac{ f}{3H_0^2}\sum_n C_n(f)P_n , 
\end{equation}
where
\begin{equation}
C_n = \frac{2n}{f^2}\int_0^z \frac{dz}{H(z) (1+z)^6}\mathtt{n}\left(\frac{2n}{(1+z)f},t(z)\right) , \label{def:Cn}
\end{equation}
with the Hubble parameter $H(z)$ and the cosmic time $t(z)$ at the redshift $z$, and the present Hubble parameter $H_0=H(0)$.
Here, $\rho_\mathrm{gw}$ is the energy density of SGWB, $\rho_\mathrm{cr}$ is the critical density, and
$\mathtt{n}\left(\ell,t\right)$ is the distribution function of the loops with length $\ell$. 
By using the dimensionless Hubble parameter $h$, we may express  
\begin{align}
\Omega_\mathrm{gw}(f)h^2 =& h^2 8\pi(G\mu)^2 \sum_n P_n \frac{2n H_0}{3f} \int_0^z \frac{H_0}{H(z)}\frac{dz}{(1+z)^6}  \frac{1}{H_0^4}\mathtt{n}\left(\frac{2nH_0}{(1+z)f},H_0t(z)\right) ,
\end{align}
where the variables of $\mathtt{n}$ are re-scaled by multiplying $H_0$ to be dimensionless.

\subsection{Loop density}

The number density of string loop $\mathtt{n}\left(\ell,t\right)$ in Eq.(\ref{def:Cn})
 encodes the dynamics of cosmic string network and is essential for the amplitude of SGWB. 
We adopt the parametrization of loop distribution of a model based
 on the Nambu-Goto (NG) simulation.
This NG based model has been widely adopted in literature as approximation, 
 because an NG model is expected to describe cosmic string dynamics for a length scale
 much larger than the the string thickness,
 though mismatch between NG and AH simulation have been reported.
For recent field theory simulations, see e.g.,
 Refs.~\cite{Blanco-Pillado:1998tyu,Olum:1998ag,Correia:2020yqg,Correia:2020gkj,Correia:2021tok,
Hindmarsh:2021mnl,Baeza-Ballesteros:2023say,Servant:2023tua,Baeza-Ballesteros:2024sny}.

In this paragraph, we note a simple review of a naive `one-scale' mode,
 only for illustrative purpose~\cite{Vilenkin:2000jqa}.
The loop energy density $\rho_{\mathrm{L}}$ follows
\begin{align}
\dot{\rho}_{\mathrm{L}}(\ell,t) = -3H \rho_{\mathrm{L}}(\ell,t)
+\sqrt{1-v_i^2}\frac{\mu}{L^4}\left(\frac{\ell}{L}\right) ,
\label{eq:Boltz:rhoL}
\end{align}
with the initial loop velocity $v_i$, the loop length $\ell$, and the correlation length scale $L$.
The second and the third term in RHS in Eq.~(\ref{eq:Boltz:rhoL}) stand for
 the dilation by cosmic expansion and the formation of loops with 
 the length $\ell$ and the tension $\mu$, respectively.
By integrating Eq.~(\ref{eq:Boltz:rhoL}), we obtain
\begin{equation}
\rho_{\mathrm{L}}(\ell,t) =  \frac{\mu\nu_\mathrm{r}}{t^{3/2}\ell^{3/2}},  \label{eq:rho_L} 
\end{equation}
where
\begin{equation}
\nu_\mathrm{r} =\frac{\sqrt{1-v_i^2}\mu}{\gamma^{5/2}}\int_{\ell/(\gamma t)}^{\infty} dx\sqrt{x}f(x) ,
\end{equation}
with $\gamma=L/t$ for the radiation dominated epoch. The loop length becomes shorter as
\begin{align}
\ell =\ell_\mathrm{i} -\Gamma G\mu(t-t_\mathrm{i}) ,
\label{eq:ell(t)}
\end{align}
by the energy loss due to gravitation radiation as Eq.~(\ref{eq:Power}) with $E=\mu \ell$.
Here, $\ell_\mathrm{i}$ is the initial length formed at time $t_\mathrm{i}$.  
Adopting the loop production function for one scale model
\begin{align}
f(x) \propto \delta(x-\alpha/\gamma), 
\end{align}
we find the string loop number density
\begin{align}
 \mathtt{n}_{\mathrm{L}}(\ell,t) = \frac{\nu_\mathrm{r}}{t^{3/2}\left(\ell+\Gamma G\mu t\right)^{5/2}}, 
\end{align}
by substituting Eq.~(\ref{eq:ell(t)}) to Eq.~(\ref{eq:rho_L}) and
 converting the energy density to number density through
\begin{align}
& \mathtt{n}_{\mathrm{L}}(\ell_\mathrm{i},t)\mu\ell_\mathrm{i} = \rho_\mathrm{L}(\ell_\mathrm{i}, t) . 
\end{align}

Several fitting formulae of the number density based on simulation has been reported
 by several groups. As in many literature, we adopt the BOS model
\begin{subequations}
\begin{align}
 \mathtt{n}_\mathrm{r,r}(l,t)
 =& \frac{0.18}{t^{3/2}(l+\Gamma G \mu t)^{5/2}}\Theta(0.1-l/t) , \\
 \mathtt{n}_\mathrm{r,m}(l,t)
 =& \frac{0.18}{(l+\Gamma G \mu t)^{5/2}}(2H_0\sqrt{\Omega})^{3/2}(1+z)^3\Theta(0.09 t_{eq}/t-\Gamma G\mu- l/t) ,
\end{align}
 \label{def:typen}%
\end{subequations}
based on simulation of Nambu-Goto strings by Blanco-Pillado et al~\cite{Blanco-Pillado:2013qja}.
Here, the subscripts \rm{r,r} stand for strings formed and GWs emitted at radiation dominated era while the subscripts \rm{r,m} stand for strings formed at radiation dominated era and GWs emitted at matter dominated epoch~\cite{Kume:2024adn}.

\subsection{string motion}

The general solution of $\mathbf{x}(\zeta,t)$ for a closed loop with the periodic boundary condition $\mathbf{x}(\zeta,t)= \mathbf{x}(\zeta+\ell,t)$ is~\cite{Vilenkin:2000jqa,Vilenkin:1986zz,Spergel:1986uu}
\begin{equation}
 \mathbf{x}(\zeta,t)=\frac{1}{2}\left(\mathbf{a}(\zeta-t)+\mathbf{b}(\zeta+t)\right), 
\label{eq:sol.x}
\end{equation}
where $\mathbf{a}$ and $\mathbf{b}$ follow 
\begin{align}
\mathbf{a}{}'^2=\mathbf{b}'{}^2=1.
\label{eq:a'b'}
\end{align}
Here, we take the coordinate $\zeta^0=t$ and work in the conformal gauge 
\begin{align}
\gamma_{01}=0,\quad  \gamma_{00}+\gamma_{11}=0, \label{eq:conf.gauge}
\end{align}
 and the prime denotes the derivative with respect to the spatial world sheet coordinate $\zeta^1$.
The period of the string motion is $\ell/2$
 because $ \mathbf{x}(\zeta+\ell/2,t+\ell/2)= \mathbf{x}(\zeta,t)$~\cite{Kibble:1982cb,Vilenkin:2000jqa}.
The point that satisfies
\begin{equation}
\mathbf{a}'=-\mathbf{b}', 
\label{eq:cusp.cond}
\end{equation}
with the prime being the derivative with respect to the variables,
 travels at the speed of light, which can be seen from Eq.~(\ref{eq:a'b'}) and 
\begin{align}
& \dot{\mathbf{x}} ^2=\frac{1}{4}\left(\mathbf{a}'(\zeta-t)-\mathbf{b}'(\zeta+t)\right)^2.
\end{align}
This is a cusp. By expanding $\mathbf{x}$ around the cusp point
 where the coordinate is taken as $(\zeta,t)=(0,0)$, we obtain
\begin{align}
\mathbf{x} =& \mathbf{x}_0 + \frac{1}{2}\left(-\mathbf{a}_0'+\mathbf{b}_0'\right)t
 + \frac{1}{2}\left(-\mathbf{a}_0''+\mathbf{b}_0''\right)t\zeta
  + \frac{1}{4}\left(\mathbf{a}_0''+\mathbf{b}_0''\right)t^2
   + \frac{1}{4}\left(\mathbf{a}_0''+\mathbf{b}_0''\right)\zeta^2 \nonumber \\
   & +\frac{1}{4} \left(\mathbf{a}_0'''+\mathbf{b}_0'''\right)t^2\zeta
   +\frac{1}{4} \left(-\mathbf{a}_0'''+\mathbf{b}_0'''\right)t\zeta^2
  + \frac{1}{12}\left(-\mathbf{a}_0'''+\mathbf{b}_0'''\right)t^3
   + \frac{1}{12}\left(\mathbf{a}_0'''+\mathbf{b}_0'''\right)\zeta^3 +\cdots ,
  \label{eq:x.series}\\
\mathbf{x}' =& \frac{1}{2}\left(-\mathbf{a}_0''+\mathbf{b}_0''\right)t
   + \frac{1}{4}\left(\mathbf{a}_0''+\mathbf{b}_0''\right)\zeta +\cdots , 
\end{align}
where the subscript $0$ stands for the value at the cusp and the linear term
 of $\zeta$ in Eq.~(\ref{eq:x.series}) is absent, $\mathbf{a}_0'+\mathbf{b}_0'=0$, due to the condition (\ref{eq:cusp.cond}) at the cusp point $(t,\zeta)=(0,0)$.
The constraint $\mathbf{x}'\!\cdot\!\dot{\mathbf{x}}=0$ gives
\begin{subequations}
\begin{align}
& \mathbf{a}_0'\!\cdot\! \mathbf{a}_0'' = \mathbf{b}_0' \!\cdot\!\mathbf{b}_0'' =0 , \\
& \mathbf{a}_0'\!\cdot\! \mathbf{a}_0''' + \mathbf{a}_0''{}^2  = 0 , \\
& \mathbf{b}_0'\!\cdot\! \mathbf{b}_0''' + \mathbf{b}_0''{}^2  = 0 .
\end{align}
\end{subequations}

\section{Particle emission from strings}
\label{sec:emissionrate}

Bearing that the broken symmetry is gauged $U(1)_{B-L}$ in mind, 
the complex scalar field in the AH model generates
 the mass of Majorana fermions, would
 be identified with right-handed neutrinos, and the gauge boson.
In this section, we calculate the energy loss rate due to
 the $Z'$ and $\psi$ emission from strings. 
In this paper, we focus on the decay through the couplings with the Higgs field. 
There are, however, additional emissions through the nontrivial configuration
 of gauge field, namely
 Aharanov-Bohm interaction~\cite{Alford:1988sj,Jones-Smith:2009adx,Chu:2010zzb}.
An extented analysis including the Aharanov-Bohm interaction will be discussed elsewhere.

\subsection{Majorana fermion from the Higgs field}

We consider emit a pair of Majorana fermion $\psi$ from cosmic strings generated
 at the symmetry breaking by a complex scalar field $\Phi$. 
The mass of $\psi$ is also generated by the VEV. The fermion part of Lagrangian is 
\begin{align}
\mathcal{L} = \frac{i}{2}\overline{\psi} \slashed{\partial}\psi - \frac{1}{2}y\overline{\psi^C}(\Phi P_R+ \Phi^* P_L)\psi ,
\label{eq:Lagrangian_for_psi}
\end{align}
 with $y$ being the real Yukawa coupling between $\Phi$ and $\psi$. Here, the superscript $C$ stands for the charge conjugation. 
The field equation is 
\begin{align}
 (i \slashed{\partial} -m) \psi &=  (y\Phi(t,\mathbf{x}) P_R+ y\Phi^*(t,\mathbf{x}) P_L -m) \psi  \simeq  -m f(\mathbf{x})\psi ,
\label{eq:EOM_of_psi}
\end{align}
 with $m=yv/\sqrt{2}$ being the mass of $\psi$ at the broken vacuum of $\Phi$ with the VEV $v$. 
Here, the function $f$ takes a value $f(\mathbf{x})=1 (0)$ at a string position $\mathbf{x}_\mathrm{st}$ (outside a string).
The interaction term may be expressed as 
\begin{align}
\mathcal{L}_\mathrm{int} 
 =& -\frac{1}{2}m f(\mathbf{x})\overline{\psi^C}\psi \nonumber \\
 \simeq & -\frac{1}{2}m\delta^2 \delta^{(2)}(\mathbf{x}-\mathbf{x}_\mathrm{st})\overline{\psi^C}\psi\nonumber \\
 =& -\frac{1}{2}m\delta^2 \int d^2\zeta \sqrt{-\gamma}\delta^{(4)}(x^{\mu}-x^{\mu}_\mathrm{st}(\zeta))\overline{\psi^C}\psi ,
\end{align}
where $\gamma$ is the induced metric on the string world sheet, $\zeta$ denotes the string world sheet coordinate, and the function $f$ is normalized as
\begin{align}
\int f(\mathbf{x})d^2x \simeq \delta^2 ,
\end{align}
 with $\delta$ being the string thickness of the order of $1/v$.
The amplitude of production of a $\psi$ pair with the momentum $\mathbf{q}_1$ and $\mathbf{q}_2$ from the string is evaluated as
\begin{align}
i\mathcal{M}=& i \int d^4x \langle \mathbf{q}_1\mathbf{q}_2 \left| \mathcal{L}_\mathrm{int}(x)\right|0\rangle \nonumber \\
 =& \frac{-i m\delta^2}{2} \overline{u}(q_1)v(q_2)\int dte^{i(q^0_1+q^0_2)t} \int^{\ell}_0 d\zeta
 \left|\mathbf{x}'\right|^2 e^{-i(\mathbf{q_1+q_2})\cdot\mathbf{x}(\zeta,t)} . 
\end{align}
The power of $\psi$ pair emission is given by
\begin{align}
P_{\psi\psi}= \frac{2}{\ell}\int q^0 dN_{\psi\psi} ,
\end{align}
with 
\begin{align}
q =& q_1+q_2, \\
dN_{\psi\psi} =& \left(\frac{m\delta^2}{2} \right)^2 2  d\Pi_{q_1} d\Pi_{q_2}\left((q_1+q_2)^2-4m^2 \right) \left|S(q_1+q_2)\right|^2 ,
\label{def:dN:psi}  
\end{align}
and 
\begin{align}
 S(q):=& \int dt \int^{\ell}_0 d\zeta
 \left|\mathbf{x}'\right|^2 e^{iq_{\mu}x^{\mu}(\zeta,t)} =\int dte^{i q^0 t} \int^{\ell}_0 d\zeta
 \left|\mathbf{x}'\right|^2 e^{-i \mathbf{q}\cdot\mathbf{x}(\zeta,t)}, 
\label{eq:S} 
\end{align}
is integrated over the string world sheet. Here, $N_{\psi\psi}$ is the number of $\psi$,
 and $d\Pi_q$ is the Lorentz invariant phase space element for the momentum $q$.
World sheet integration is followed by Ref.~\cite{Srednicki:1986xg} and
 some detail is summarized in App.~\ref{sec:world sheet integration}.
Our Eq.~(\ref{def:dN:psi}) agrees with Eq.~(B.34) in Ref.~\cite{Long:2014mxa} except for the sign in front of $4m^2$. 
The origin of the difference is clear. 
This minus sign is due to the Majorana nature, while it is plus for a Dirac fermion~\cite{Long:2014mxa}.  
This mass dependence does not play an important role nevertheless,
 because the particle emission from strings are UV dominated.
The phase space integration proceeds as  
\begin{align}
\frac{dN_{\psi\psi}}{dq_+} =& \left(\frac{m\delta^2}{2} \right)^2\frac{ 2 \left( \mathbf{a}_0''+\mathbf{b}_0'' \right)^2}{(2\pi)^4} \int \frac{dq_- d |\mathbf{q}|^2}{8}\frac{ \left(q_+^2-|\mathbf{q}|^2-4m^2 \right)}{4|\mathbf{q}|^2\left(\mathbf{q}\!\cdot\!\left( \mathbf{a}_0'''+\mathbf{b}_0''' \right)\right)^{4/3}}  \nonumber \\
\simeq & \left(\frac{m\delta^2}{2} \right)^2\frac{ 2 \left( \mathbf{a}_0''+\mathbf{b}_0'' \right)^2}{(2\pi)^4 \left( \mathbf{a}_0'''+\mathbf{b}_0''' \right)^{4/3}} \int \frac{  dq_- }{8} d |\mathbf{q}|^2\frac{ \left(q_+^2-|\mathbf{q}|^2-4m^2 \right)}
{4 \left(|\mathbf{q}|^2\right)^{5/3}  }     \nonumber \\
\simeq &  \left(\frac{m\delta^2}{2} \right)^2\frac{ 3 \ell^{2/3}  m^{5/3}}{16 (2\pi)^4} \mathcal{I}_{\psi}(q_+) ,  \label{eq:intPpsi}\\
\mathcal{I}_{\psi}(q_+) =& \int d\frac{q_-}{m} \left[-\frac{\left((q_+/m)^2-4 \right)}{2\left(|\mathbf{q}|^2/m^2 \right)^{2/3}}-\left(\frac{|\mathbf{q}|^2}{m^2} \right)^{1/3}    \right]^{|\mathbf{q}|^2_{\mathrm{upper}}}_{|\mathbf{q}|^2_{\mathrm{lower}}} . \label{eq:calIpsi} 
\end{align}
This integration range is derived in App.~\ref{sec:phase space integration}.
Here, we approximate
\begin{subequations}
\begin{align}
& \mathbf{q}\!\cdot\!\left( \mathbf{a}_0'''+\mathbf{b}_0''' \right)\simeq |\mathbf{q}|| \mathbf{a}_0'''+\mathbf{b}_0''' | ,\\
& |\mathbf{a}_0''+\mathbf{b}_0'' |\simeq \ell^{-1}, \qquad |\mathbf{a}_0'''+\mathbf{b}_0''' |\simeq \ell^{-2},
\end{align}
\label{eq:approx}%
\end{subequations}
in Eq.~(\ref{eq:intPpsi}), and
 the integration range in Eq.~(\ref{eq:calIpsi}) is also approximated
 by using Eq.~(\ref{eq:q_bound2}) instead of Eq.~(\ref{eq:q_bound}),
 because the threshold is not important.
This function, which means $dN_{\psi\psi}/dq_+$ as well, is asymptotically 
\begin{align}
\mathcal{I}_{\psi}(q_+) \simeq 0.05 \left(\frac{q_+}{m}\right)^{2/3} \qquad \mathrm{for} \qquad q_+ \gg m. 
\end{align}
We obtain
\begin{align}
P_{\psi\psi} 
= \frac{2}{\ell} \int dq_+ q_+ \frac{dN_{\psi\psi}}{d q_+} \simeq \frac{1}{\ell}\left(\frac{m\delta^2}{2} \right)^2\frac{ 9 \ell^{2/3} m^{11/3}}{1280 (2\pi)^4 } \left(\frac{q_\mathrm{UV}}{m}\right)^{8/3}
\simeq \left(\frac{m \delta}{2} \right)^2\frac{9m \delta^{4/3} }{1280 (2\pi)^4 \ell^{1/3} }  ,
\end{align}
where the upper bound of $q_+$ integration is taken to be $q_\mathrm{UV} \simeq 1/\delta$ as the UV cutoff.

\subsection{Gauge boson radiation}

If $\Phi$ is a Higgs field for a $U(1)$ gauge symmetry as in the $U(1)_{B-L}$ model,
 in the outer region of cosmic strings 
 the associated gauge boson acquires the mass $q_{\Phi}^2 g^2 v^2$,
 where $g$ is the gauge coupling, and $q_{\Phi}$ is the charge of $\Phi$.
The field equation of $Z'^{\mu}$ is
\begin{align} 
&\partial^{\nu}Z'_{\mu\nu}+2 (q_{\Phi}g)^2 |\Phi|^2Z'_{\mu} = J_{\mu} ,
 \label{eq:eom:X} \\
& J_{\mu}:= i q_{\Phi}g \left(\Phi^* \partial_{\mu} \Phi -\Phi \partial_{\mu}\Phi^* \right),
\end{align}
where $Z'^{\mu\nu}$ is the field strength, and $J^{\mu}$ is the source term.
The production amplitude of $Z'$ pair with the momentum $\mathbf{q}_1$ and $\mathbf{q}_2$ from a string is evaluated as
\begin{align}
i\mathcal{M}=& i \int d^4x \langle \mathbf{q}_1\mathbf{q}_2 \left| \mathcal{L}_\mathrm{int}(x)\right|0\rangle 
 = \varepsilon^*_{\mu}(q_1)\frac{-i m_{Z'}^2\delta^2}{2} \varepsilon^*{}^{\mu}(q_2) S(q).
\end{align}
We find
\begin{align}
dN_{Z'Z'} = \left(\frac{m_{Z'}^2\delta^2}{2}\right)^2
 d\Pi_{q_1} d\Pi_{q_2}\left(2+\left(\frac{(q_1+q_2)^2}{2m_{Z'}^2}-1\right)^2\right)\left|S(q)\right|^2 ,
\end{align}
where the part of $(q_1+q_2)^2/m_{Z'}^2$ inside brackets comes from the longitudinal mode of the massive $Z'$ boson.
This causes sizable enhancement for the relativistic products and its blue tilted spectrum.
By using Eq.~(\ref{eq:approx}), we obtain  
\begin{align}
P_{Z'Z'} = \frac{2}{\ell}\int dq_+ q_+ \frac{d N_{Z'Z'}}{dq_+} 
\simeq \left(\frac{m_{Z'}^2\delta^2}{2}\right)^2
\frac{ 81  \ell^{2/3} }{224(2\pi)^4  }\frac{q_\mathrm{UV}^{14/3}}{\ell m_{Z'}^3} .
\end{align}

\section{SGWB spectrum}
\label{sec:spectrum}

\subsection{Energy loss of strings}

\begin{figure}[htbp]
    \centering
    \includegraphics[width=0.7\textwidth]{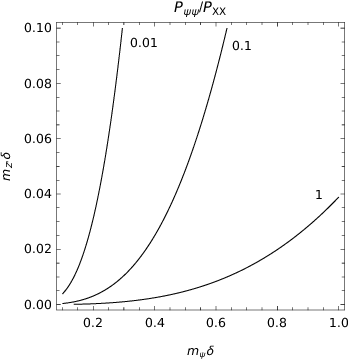}
\caption{The ratio of the emission power $P_{\psi\psi}/P_{Z'Z'}$.}
\label{Fig:ratio}
\end{figure}

The power of the particle emission from strings is read as
\begin{align}
P_\mathrm{par} =& P_{\psi\psi} + P_{Z'Z'}, \\
P_{\psi\psi} \simeq & \frac{m^3 \ell}{4} \frac{9}{1280 (2\pi)^4}
 \left(\frac{\delta}{\ell}\right)^{4/3}  ,\\
P_{Z'Z'} \simeq & \frac{m_{Z'}}{4\ell}
\frac{ 81 }{448(2\pi)^4} \left(\frac{\ell}{\delta}\right)^{2/3} ,
\end{align}
from the calculations in the previous section.
Here, we note that this power shows differnet parameter dependence on $\ell$ and $\delta$
 as $P_{Z'Z'} \propto (\delta/\ell)^{1/3}$ which somewhat differs from
 the formula $P_{Z'Z'} \propto (\ell_\mathrm{cr}/\ell)^{1/2}$~\cite{Auclair:2019jip} implied by NG string simulations~\cite{Blanco-Pillado:1998tyu,Olum:1998ag}.
The ratio of the power is found as 
\begin{equation}
R=\frac{P_{\psi\psi}}{P_{Z'Z'}} \simeq \frac{7}{180}\frac{\delta^2 m^3}{m_{Z'}},
\end{equation}
which is independent from $\ell$, and shown in Fig.~\ref{Fig:ratio}.
Unless the spectrum is of $m_{Z'} \ll m$, the radiation of gauge bosons are much efficient than that of fermions.

By taking this energy loss into account, Eq.~(\ref{eq:Power}) is modified as
\begin{align} 
\dot{E } =& -\Gamma G\mu^2 - P_\mathrm{par} , 
\label{def:modPower}  \\
P_\mathrm{par} = & \frac{81}{1792(2\pi)^4}\frac{m_{Z'}}{\ell^{1/3} \delta^{2/3}}\left(1+R \right) .
\end{align}
Since $P_\mathrm{par}$ is proportional to $\ell^{-1/3}$,
Eq.~(\ref{def:modPower}) is rewritten as 
\begin{align} 
\dot{E} =& -\Gamma G\mu^2 - \left(\frac{\ell_\mathrm{cr}}{\ell}\right)^{1/3}\left.P_\mathrm{par}\right|_{\ell=\ell_\mathrm{cr}} ,
\end{align}
where the critical length 
\begin{align} 
\ell_\mathrm{cr} :=& \left(\frac{\pi}{\mu \delta^2}\right)^3  \left(\frac{81}{1792\pi (2\pi)^4 }\frac{1}{\Gamma G\mu}\right)^3 
  \left( \left(1+R \right) m_{Z'} \delta \right)^3 \delta   \label{def:lcr}  \nonumber \\
 =& \left(\frac{\pi}{\mu \delta^2}\right)^{5/2}\left(\frac{81}{1792\pi (2\pi)^4 }\frac{1}{\Gamma G\mu}\right)^3 
 \frac{ \left( \left(1+R \right) m_{Z'} \delta \right)^3 }{\sqrt{8 G\mu}}\frac{1}{M_P}
\end{align}
 is defined from 
\begin{equation}
\Gamma G\mu^2 = \left.P_\mathrm{par}\right|_{\ell=\ell_\mathrm{cr}}.
\label{def:Ppar}
\end{equation}
Here, the normalization of the second factor of RHS in Eq.~(\ref{def:lcr}) is motivated
 by the relation $\mu = \pi v^2$ in the AH string with
 the critical coupling~\cite{Vilenkin:2000jqa}.

\begin{figure}[htbp]
    \centering
    \includegraphics[width=0.7\textwidth]{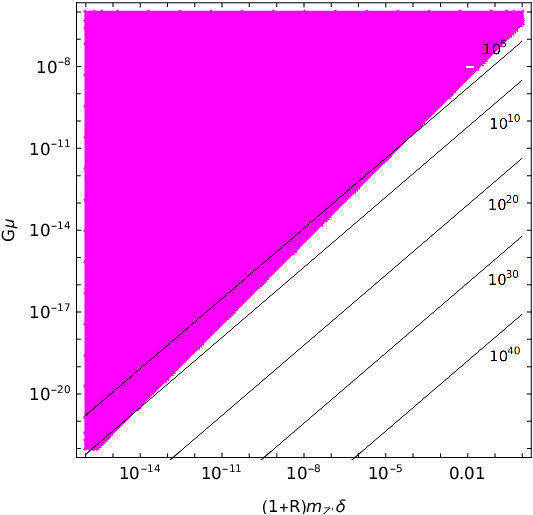}
\caption{The critical length $\ell_\mathrm{cr}$ with $\mu\delta^2=\pi$ for various $G\mu$.}
\label{Fig:lcr}
\end{figure}

The energy emission is dominated by the gravitational radiation for longer strings $\ell > \ell_\mathrm{cr}$
 and by particle production for shorter strings $\ell < \ell_\mathrm{cr}$.
Figure~\ref{Fig:lcr} shows the contours of various $\ell_\mathrm{cr}$
 in the coupling versus $G\mu$ plain.
The numbers associated with contours are the length in the unit of Planck length.
Magenta shaded region stand that the critical length becomes shorter than the string thickness
 as $\ell_\mathrm{cr} \leq \delta$, where our formulation based on the Nambu-Goto approximation
 is not verified.
While the particle emission cannot be effective for $G\mu \gtrsim 10^{-7}$.
The lower $G\mu$, the critical length $\ell_\mathrm{cr}$ becomes longer
 for a fixed interaction strength.

The solution of Eq.~(\ref{def:modPower}) is 
\begin{subequations}
\begin{align} 
\ell(t) & \simeq \ell_\mathrm{i} - \Gamma G\mu (t-t_\mathrm{i})  & \mathrm{for} \qquad \ell > \ell_\mathrm{cr} , \label{eq:longersol} \\
\ell(t) & \simeq \ell_\mathrm{cr}\left(\frac{4\Gamma G\mu}{3 \ell_\mathrm{cr}} (t_\mathrm{cr}-t)+1 \right)^{3/4}
  & \mathrm{for} \qquad \ell < \ell_\mathrm{cr} ,\label{eq:shortersol}
\end{align}
\end{subequations}
with
\begin{equation} 
t_\mathrm{cr} \simeq \frac{\ell_\mathrm{i}-\ell_\mathrm{cr}}{\Gamma G\mu } ,
\end{equation}
which is defined by $\ell(t_\mathrm{cr})=\ell_\mathrm{cr}$. Here Eq.~(\ref{def:Ppar}) is used. 
From  Eq.~(\ref{eq:ell(t)}) and Eq.~(\ref{eq:shortersol}), we obtain 
\begin{subequations}
\begin{align} 
\ell_\mathrm{i} &\simeq \Gamma G\mu  t+ \ell     & \mathrm{for} \qquad t>t_\mathrm{cr}, \\
\ell_\mathrm{i} &\simeq \Gamma G\mu  t+ \ell_\mathrm{cr}\left(\frac{3 }{4}\left(\frac{\ell(t)}{\ell_\mathrm{cr}}\right)^{4/3}+\frac{1}{4} \right)
& \mathrm{for} \qquad t>t_\mathrm{cr}  .
\end{align}
\end{subequations}
This can be well fitted by 
\begin{align} 
\ell_\mathrm{i} = \Gamma G\mu  t+ \ell_\mathrm{cr}\left(\frac{3 }{4}\left(\frac{\ell(t)}{\ell_\mathrm{cr}}\right)^{4/3}+\frac{1}{4} \right)e^{-2\ell/\ell_\mathrm{cr}} + \frac{\ell}{1+e^{-2\ell/\ell_\mathrm{cr}}} ,
\end{align}
which is replaced with $\ell+\Gamma G \mu t$ in the formula (\ref{def:typen})
 and used in our numerical calculation.

\subsection{SGWB Spectrum}

\begin{figure}[htbp]
    \centering
    \includegraphics[width=0.9\textwidth]{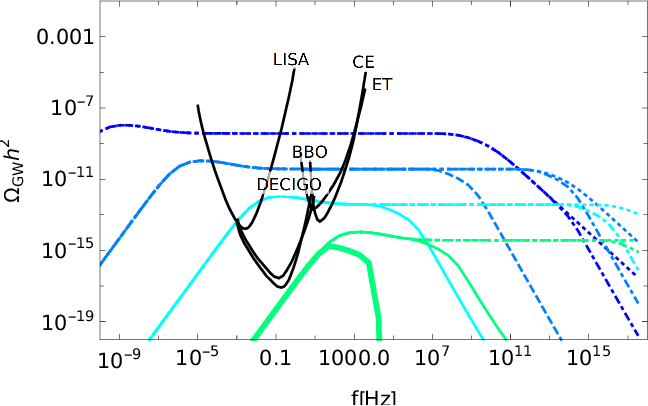}
\caption{The GW spectrum. $G\mu =10^{-9}$ (blue), $10^{-13}$ (sky blue), $10^{-17}$ (cyan) and $10^{-21}$ (green).}
\label{Fig:spectrum}
\end{figure}

Figure~\ref{Fig:spectrum} displays the generated SGWB spectrum for various $G\mu$ and
 various $\ell_\mathrm{cr}$.
In the following, we fix $\mu\delta^2=\pi$ and take $G\mu$ and $\ell_\mathrm{cr}$ as input free parameters. 
Dotted curves corresponds the case without particle emission, in other words $\ell_\mathrm{cr}=0$.
This is nothing but the known prediction of SGWB from cosmic strings found in the previous works~\cite{Caldwell:1991jj}.
This shows the characteristic feature of spectrum:
 a drop at low frequencies and a flat plateau at high frequencies. The SGWB in the damped region was emitted during the matter dominated era,
 while SGWB with the frequency at the flat spectrum region were emitted at radiation dominated era. 
The cutoff at the very high frequency corresponds to SGWB emitted from the smallest loops
 at the earliest time~\cite{Caldwell:1991jj,Gouttenoire:2019kij,Sousa:2020sxs}. 

Dot-dashed, dashed, solid and thick solid curves correspond to $\ell_\mathrm{cr}=10^{10}, 10^{20}, 10^{30}$
 and $10^{40}$ in the Planck unit, respectively.
The effect of particle radiation on the SGWB spectrum appears as cutoff at high frequencies,
because the energy loss due to the particle emission is effective
 only for $\ell \leq \ell_\mathrm{cr}$.
For $G\mu=10^{-9}$, the flat spectrum does not change and
 only the power of the damp at the very high frequency becomes larger.
For smaller values of $G\mu$, damping of the flat spectrum appears and
 its transition frequency depends on $\ell_\mathrm{cr}$.
The $\ell_\mathrm{cr}$ longer, the lower the transition frequency. 
In our benchmark with $G\mu=10^{-21}$ which corresponds to $v=\mathcal{O}(10^8)$ GeV,
 the flat region of the SGWB spectrum disappears for $\ell_\mathrm{cr}M_P > 10^{40}$.
For $(G\mu, \ell_\mathrm{cr}M_P)=(10^{-21}, 10^{50})$,
 almost all energy of cosmic strings are radiated by the particle emission
 and little SGWB are produced, hence it is not displayed. 
This parameter set corresponds to the gauge coupling about $\mathcal{O}(0.1)$.
For such $v$ and the gauge coupling, detectable SGWB could be generated
 by the phase transition rather than cosmic strings
 if it is strong first order~\cite{Okada:2018xdh,Hasegawa:2019amx,Okada:2020vvb}.

\section{Summary}
\label{sec:summary}

We have studied the effect of particle radiation
 on the SGWB spectrum generated
 from cosmic strings in an Abelian-Higgs system.
We have formulated the calculation
 for the emission of gauge bosons and Majorana fermions
 whose masses are generated by the Higgs field forming cosmic strings.

We have found that the energy loss is more effective by the gauge boson emission
 than by the Majorana fermion emission in the wide range of parameter space. 
The effect on the SGWB spectrum appears only in the high frequency region,
 because the energy loss of strings by the particle radiation is effective
 only for shorter stings than $\ell_\mathrm{cr}$.
Although the suppression of the SGWB spectrum significantly narrows the flat spectrum region
 for $G\mu \ll 10^{-9}$, its frequency at which th edamp begins in general
 too high to observe for planed interferometer experiments.

%
\section*{Acknowledgments}

This work is supported in part by the U.S. DOE Grants No. DE-SC0012447 and DE-
SC0026347 (N.O.) and KAKENHI Grants No. JP23K03402 (O.S.).

\appendix

\section{world sheet integration}
\label{sec:world sheet integration}

By substituting Eq.~(\ref{eq:x.series}), the phase of the exponential part in Eq.~(\ref{eq:S})
 around the cusp $(t,\zeta)=(0,0)$ is expressed as
\begin{align} 
\psi =& q^0 t-\mathbf{q}\!\cdot\!\mathbf{x}(\zeta,t) \nonumber \\
\simeq & -\mathbf{q}\!\cdot\!\left(\frac{1}{4} \left(\mathbf{a}_0'''+\mathbf{b}_0'''\right)t^2\zeta
   +\frac{1}{4} \left(-\mathbf{a}_0'''+\mathbf{b}_0'''\right)t\zeta^2
  + \frac{1}{12}\left(-\mathbf{a}_0'''+\mathbf{b}_0'''\right)t^3
   + \frac{1}{12}\left(\mathbf{a}_0'''+\mathbf{b}_0'''\right)\zeta^3 \right)\nonumber \\
 & +q^0 t-\left( \mathbf{q}_{\parallel}\!\cdot\!\dot{\mathbf{x}}_0 t
 + \frac{1}{2}\mathbf{q}_{\perp}\!\cdot\!\left(-\mathbf{a}_0''+\mathbf{b}_0''\right)t\zeta
  + \frac{1}{4}\mathbf{q}_{\perp}\!\cdot\!\left(\mathbf{a}_0''+\mathbf{b}_0''\right)t^2
   + \frac{1}{4}\mathbf{q}_{\perp}\!\cdot\!\left(\mathbf{a}_0''+\mathbf{b}_0''\right)\zeta^2 \right),
\label{eq:psi}
\end{align}
with $\mathbf{x}_0=0$, $\mathbf{q} = \mathbf{q}_{\parallel}+\mathbf{q}_{\perp}$ and $\mathbf{q}_{\perp}\!\cdot\!\dot{\mathbf{x}}_0 =0$,
The dominant contribution comes from the period near the cusp for before rapid oscillation. 
$\psi \lesssim 1$ corresponds to the integration range 
\begin{align}
\zeta, t \lesssim \mathrm{min}[ \left(\mathbf{q}\!\cdot\!\left( \mathbf{a}_0'''\pm\mathbf{b}_0''' \right)\right)^{-1/3}, \ell]  , 
\label{eq:lange_tzeta}
\end{align}
which indicates that the total momentum of emitted fermion pair is bounded as
\begin{align}
|\mathbf{q}|_{\mathrm{min}} :=
\frac{1}{|\mathbf{a}_0'''\pm\mathbf{b}_0'''|\ell^3} \lesssim  |\mathbf{q}| .
 \label{eq:worldsheet:q_bound}
\end{align}
This means that the de Broglie length of emitted quanta is smaller than the loop length.
The range of the relative angle between $\mathbf{q}$ and $\dot{\mathbf{x}}_0$
 denoted by $\varphi$ turns out to be
\begin{align}
\varphi \lesssim \varphi_{\mathrm{max}} :=
\frac{ \left(\mathbf{q}\!\cdot\!\left( \mathbf{a}_0'''\pm\mathbf{b}_0''' \right)\right)^{2/3}}{|\mathbf{q}||\mathbf{a}_0''\pm\mathbf{b}_0''|} ,
\end{align}
from $\mathbf{q}_{\perp}$ dependent parts in Eq.~(\ref{eq:psi}).

For a $\mathbf{q}$ satisfying the condition (\ref{eq:worldsheet:q_bound}), we find
\begin{align} 
S \simeq & \int_0^{\ell/2} dt\int^{\ell}_0 d\zeta 
 \left|\left(-\mathbf{a}_0''+\mathbf{b}_0''\right)t
   + \left(\mathbf{a}_0''+\mathbf{b}_0''\right)\zeta \right|^2 \nonumber \\
&   \exp\left[i \left\{ q^0 t-\mathbf{q}\!\cdot\!\left(\dot{\mathbf{x}}_0t +\frac{\left(\mathbf{a}_0'''+\mathbf{b}_0'''\right)}{12} \left( 3t^2\zeta + \zeta^3 \right)
   +\frac{\left(-\mathbf{a}_0'''+\mathbf{b}_0'''\right)}{12} \left(3t\zeta^2 + t^3\right) \right)\right\} \right]  \nonumber \\  
 \simeq &  \left| \mathbf{a}_0''+\mathbf{b}_0'' \right|^2 \left(\mathbf{q}\!\cdot\!\left( \mathbf{a}_0'''+\mathbf{b}_0''' \right)\right)^{-4/3} ,
\end{align}
where we have used Eqs.~(\ref{eq:lange_tzeta}).

\section{phase space integration}
\label{sec:phase space integration}

The phase space volume element is calculated as 
\begin{align} 
 d\Pi_{q_1} d\Pi_{q_2} = \frac{|\mathbf{q}_1|dq_1^0}{(2\pi)^2} \frac{|\mathbf{q}_2|dq_2^0}{(2\pi)^2}\frac{d\cos\theta}{2}\frac{d\cos\varphi}{2} = \frac{1}{(2\pi)^4} \frac{d\cos\varphi}{2}\frac{dq_+ dq_- d |\mathbf{q}|^2}{8}  ,
\end{align}
by changing integration variables $q_1^0, q_2^0$ and $\theta$,
 which is the angle between $\mathbf{q}_1$ and $\mathbf{q}_2$, 
 with the integration range are $q_1^0 \geq m, q_2^0\geq m$ and $|\cos\theta| \leq 1$, to
\begin{subequations}
\begin{align}
 q_+ =& q_1^0+q_2^0 , \\
 q_- =& q_1^0-q_2^0 , \\
 |\mathbf{q}|^2 =& |\mathbf{q}_1+\mathbf{q}_2|^2
= |\mathbf{q}_1|^2+ |\mathbf{q}_2|^2 +2|\mathbf{q}_1||\mathbf{q}_2| \cos\theta .
\end{align}
\end{subequations}
The integration range are converted to $2m < q_+ < \infty$, $q_+ -2m <q_-<q_+ +2m$ and 
 $|\mathbf{q}|^2_{\mathrm{lower}}<|\mathbf{q}|^2<|\mathbf{q}|^2_\mathrm{upper}$ with 
\begin{align}
|\mathbf{q}|^2_{ \mathrm{upper}, \mathrm{lower}}&
 = \frac{1}{2}\left(q_-^2+q_+^2-4m^2\pm\sqrt{(q_-^2+q_+^2-4m^2)^2-4 q_-^2 q_+^2}\right).
 \label{eq:q_bound}
\end{align}
As long as a momentum is away from the threshold, Eq.~(\ref{eq:q_bound}) may be simplified as
\begin{subequations}
\begin{align}
|\mathbf{q}|^2_{\mathrm{upper}}& = q_-^2+q_+^2-4m^2, \\
|\mathbf{q}|^2_{\mathrm{lower}}& = \frac{q_-^2 q_+^2}{q_-^2+q_+^2-4m^2}.
\end{align}
 \label{eq:q_bound2}
\end{subequations}
%



\end{document}